\documentclass[pra,tightenlines,showpacs,nofootinbib,,superscriptaddress]{revtex4}
\usepackage{dcolumn,bm,graphicx,amsmath,amsfonts,amssymb}
\usepackage
{hyperref}

\begin{document}
\title{Including of triple excitations in the relativistic
coupled-cluster formalism and calculation of Na properties. }

\author{Sergey G. Porsev}
\affiliation {Department of Physics, University of Nevada, Reno,
Nevada 89557, USA}
\affiliation{Petersburg Nuclear Physics
Institute, Gatchina, Leningrad district, 188300, Russia}
\author{Andrei Derevianko}
\affiliation {Department of Physics, University of Nevada, Reno,
Nevada 89557, USA}

\date{\today}

\begin{abstract}
A practical high-accuracy relativistic method of atomic structure
calculations for univalent atoms is presented. The method is rooted
in the coupled-cluster formalism and includes non-perturbative
treatment of single and double excitations from the core and single,
double and triple excitations involving valence electron. Triple
excitations of core electrons are included in the fourth-order of
many-body perturbation theory. In addition, contributions from the
disconnected excitations are incorporated. Evaluation of matrix
elements includes all-order dressing of lines and vertices of the
diagrams. The resulting formalism for matrix elements is complete
through the fourth order and sums certain chains of diagrams to all
orders. With the developed method we compute removal energies,
magnetic-dipole hyperfine-structure constants $A$ and
electric-dipole amplitudes. We find that the removal energies are
reproduced within 0.01-0.03\% and the hyperfine constants of the
$3s_{1/2}$ and $3p_{1/2}$ states with a better than 0.1\% accuracy.
The computed dipole amplitudes for the principal
$3s_{1/2}-3p_{1/2;3/2}$ transitions  are in an agreement with
0.05\%-accurate experimental data.
\end{abstract}


\maketitle

\section{Introduction}
\label{Sec:Intro}

This work is aimed at designing a practical {\em ab initio}
atomic-structure method capable of reaching accuracy at the level of
0.1\% for properties of heavy univalent many-electron atomic
systems.
The improved accuracy is required, for example, for a  refined
interpretation of atomic parity violation (APV) with atomic
Cs~\cite{Khr91,BouBou97,WooBenCho97} and planned experiment with
Ba$^{+}$~\cite{KoeSchNag03}. At present namely the accuracy of
solving the basic correlation problem is the limiting factor in the
APV probe of ``new physics'' beyond the standard model of elementary
particles. In addition,  it is anticipated that the improved
accuracy would unmask so far untested contributions from quantum
electrodynamics (QED) in heavy neutral many-electron
systems~\cite{SapChe03}.

Here we report developing a many-body approach based on the
coupled-cluster (CC) formalism~\cite{CoeKum60,Ciz66}. In the CC
formalism the many-body contributions to wave function are lumped
into a hierarchy of multiple (single, double, ...) particle-hole
excitations from the lowest-order state. Due to  a computational
complexity, previous relativistic CC-type
calculations~\cite{BluJohLiu89,BluJohSap91,EliKalIsh94,AvgBec98,SafDerJoh98,SafJohDer99}
for univalent atoms were limited to single-\ and double excitations.
Triple excitations were treated only in an approximate
semi-perturbative fashion
~\cite{BluJohLiu89,BluJohSap91,SafDerJoh98,SafJohDer99,GopMerMaj01,ChaSahDas03}.
Compared to these previous calculations, here we fully include
valence triple excitations in the CC formulation; we will designate
our approximation as CCSDvT method. Further, compared to
calculations by Notre Dame group, here we also incorporate a subset
of so-called disconnected excitations (non-linear CC terms). For
sodium atom, such non-linear CC terms were previously included in
Ref.~\cite{EliKalIsh94} and in non-relativistic
calculations~\cite{SalYnn91}. Finally, in calculations of matrix
elements we include CC-dressing of lines and
vertices~\cite{DerPor05} and we also directly compute complementary
fourth-order diagrams (mainly due to core triple excitations). The
resulting formalism for matrix elements is complete through the
fourth-order of many-body perturbation theory (MBPT) and also
subsumes certain chains of diagrams to all orders.


As a first application of our method, we  carry out numerical
calculations for atom of sodium. Sodium (11 electrons) has an
electronic structure similar to cesium (55 electrons), but it is not
as demanding computationally. By computing properties of Na atom we
observe that  a simultaneous treatment of triple and disconnected
quadruple excitations is important for improving theoretical
accuracy, as the two effects tend to partially cancel each other. We
compute removal energies, magnetic-dipole hyperfine-structure (HFS)
constants $A$ and electric-dipole amplitudes for the principal
$3s_{1/2}-3p_{j}$ transitions. We find that the removal energies are
reproduced within 0.01-0.03\%  and the HFS constants of the $3s$ and
$3p_{1/2}$ states with a better than 0.1\% accuracy. The computed
dipole amplitudes are in a perfect agreement with the
0.05\%-accurate experimental data. However, our result for the HFS
constant of the $3p_{3/2}$ state disagrees with the most accurate
experimental values~\cite{YeiSieHav93,GanKarMar98} by ~1\%, while
agreeing with less accurate
measurements~\cite{KriKusGau77,AriIngVio77}.

The paper is organized as follows. First we discuss generalities of
the coupled-cluster formalism and many-body perturbation theory in
Sec.~\ref{Sec:Generalities}. Explicit CCSDvT equations and
analytical expressions for energies, matrix elements, and
normalization corrections are presented in
Section~\ref{Sec:Formalism}. In Sec.~\ref{Sec:Results} we tabulate
and analyze the results of numerical calculations of properties of
sodium atom. Finally, we draw conclusions in Sec.~\ref{Sec:concl}.
Unless specified otherwise, atomic units
$|e|=\hbar=m_e=4\pi\varepsilon_0\equiv 1$ are used throughout.

\section{Generalities}
\label{Sec:Generalities}
In this Section we recapitulate relevant
formulas and ideas of atomic many-body perturbation theory (MBPT)
and the coupled-cluster formalism for systems with one valence
electron outside the closed-shell core.
\subsection{Atomic Hamiltonian and conventions}
The Hamiltonian of an atomic system may be represented as
\begin{equation}
H =
\left( \sum_{i}\,h_{\rm nuc}(\mathbf{r}_i) + \sum_{i}\, U_{\rm DHF}(\mathbf{r}_i) \right)
+
\left( \frac{1}{2}\sum_{i \neq j}\frac{1}{r_{ij}}
-\sum_{i}\,U_{\rm DHF}(\mathbf{r}_i) \right) \,,
\label{Eqn_Htot}
\end{equation}
where $h_{\rm nuc}$ is the Dirac Hamiltonian including  kinetic
energy of electron and its interaction with the nucleus, $U_{\rm
DHF}$ is the Dirac-Hartree-Fock (DHF) potential, and the last term
represents the residual Coulomb interaction between electrons. To
reduce the number of MBPT diagrams, we employ frozen-core (or
$V^{N-1}$) DHF potential~\cite{Kel69}. The single-particle orbitals
$\varphi_i$ and energies  $\varepsilon_i$ are found from the set of
DHF equations,
\begin{equation}
\left( h_{\rm nuc} + U_{\rm DHF}  \right) \varphi_i = \varepsilon_i \varphi_i \, .
\end{equation}

The Hamiltonian in the second quantization  reads (omitting common
energy offset)
\begin{equation}
 H =  H_0 + G = \sum_{i} \varepsilon_i N[a_i^\dagger a_i] +
 \frac{1}{2} \sum_{ijkl} g_{ijkl} N[a^\dagger_i a^\dagger_j a_l a_k ] \, ,
 \label{Eq:SecQuantH}
\end{equation}
where operators $a_{i}$ and $a_{i}^\dagger$ are annihilation and
creation operators, and $N[\cdots]$ stands for a normal product of
operators with respect to the core quasi-vacuum state $|0_c\rangle$.
Labels $i, j,k$ and $l$ range over all possible single-particle
orbitals. In the following we will employ a labeling convention
where letters $a,b,c$ are reserved for core orbitals, indices
$m,n,r,s$ label virtual states, and letters $v$ and $w$ designate
valence orbitals. In this convention valence orbitals are classified
as the virtual orbitals. In Eq.~(\ref{Eq:SecQuantH}), the quantities
$g_{ijkl}$ are two-body Coulomb matrix elements
\begin{equation}
g_{ijkl}=\int d^{3}\mathbf{r}\int d^{3}\mathbf{r}'
\varphi_{i}^{\dagger}(\mathbf{r})\varphi
_{j}^{\dagger}(\mathbf{r}^{\prime})\frac{1}{\left\vert \mathbf{r}%
-\mathbf{r}^{\prime}\right\vert }\varphi_{k}(\mathbf{r})\varphi_{l}(\mathbf{r}%
^{\prime})\,.
\label{Eq:g_ijkl}
\end{equation}
Notice the absence of the one-body contribution of $G$ in the
second-quantized Hamiltonian, Eq.~(\ref{Eq:SecQuantH}); this
simplifying feature is due to the employed  $V^{N-1}$ approximation
and leads to a greatly reduced number of terms in the CC equations.

In MBPT the first part of the Hamiltonian~(\ref{Eq:SecQuantH})  is
treated as the lowest-order Hamiltonian $H_0$ and the residual
Coulomb interaction $G$ as a perturbation.
In the lowest order the atomic wave function with the valence
electron in an orbital $v$ reads $|\Psi_v^{(0)} \rangle =
a^\dagger_v | 0_c \rangle$. Further the wave operator $\Omega$ is
introduced; it promotes  this lowest-order state to the exact
many-body wave function
\begin{equation}
|\Psi_v\rangle = \Omega\, |\Psi_v^{(0)}\rangle.
\end{equation}
In the conventional order-by-order MBPT, a perturbative expansion
for operator $\Omega$ is built in powers of residual interaction $G$
resulting in  a hierarchy of approximations for correlated energies
and wave-functions.

\subsection{Coupled-cluster method}
One of the mainstays of practical application of MBPT is an
assumption of convergence of series in powers of the perturbing
interaction. Sometimes the convergence is poor and then one sums
certain classes of diagrams  to ``all orders'' using iterative
techniques. The coupled-cluster  formalism is one of the most
popular all-order methods. The key point of the CC method is the
introduction of an exponential ansatz for the wave
operator~\cite{LinMor86}
\begin{equation}
 \Omega = N[ \exp(K)] = 1 + K + \frac{1}{2!} N[K^2] + \ldots \, ,
 \label{Eq:CCOmega}
\end{equation}
where the cluster operator $K$ is expressed in terms of connected
diagrams of the wave operator $\Omega$. The operator $K$ is
naturally broken into cluster operators $\left(K\right)_n$ combining
$n$ simultaneous excitations of core and valence electrons from the
reference state $|\Psi_v^{(0)}\rangle$ to all orders of MBPT
\begin{equation}
K=\sum_n^\textrm{total number of electrons} \left(K\right)_n = S + D + T + \cdots,
\end{equation}
i.e., $K$ is separated into singles
($S\equiv\left(K\right)_1$), doubles
($D\equiv\left(K\right)_2$), triples
($T\equiv\left(K\right)_3$), etc. For the univalent
systems we further separate the cluster operators
into two, core and valence, classes
\begin{equation}
\left(K\right)_n = \left(K_c\right)_{n} + \left(K_v\right)_{n} \,.
\end{equation}
Clusters $\left(K_c\right)_{n}$ involve excitation from the core
orbitals only, while $\left(K_v\right)_{n}$ describe simultaneous
excitations of the core and valence electrons. Then $S=S_c + S_v$,
$D=D_c+D_v$, etc.

A set of coupled equations for the cluster operators
$\left(K\right)_n$ may be found  from the Bloch
equation~\cite{LinMor86} specialized for univalent
systems~\cite{DerEmm02}
\begin{eqnarray}
\left(  \varepsilon_{v}  -H_{0}\right)
\left(K_c\right)_{n} &=&
\left\{  Q\, G \,\Omega\right\}_{\mathrm{connected},n}\, , \nonumber\\
\left(  \varepsilon_{v}+ \delta E_v   -H_{0}\right)
\left(K_v\right)_{n} &=&
\left\{  Q\, G \,\Omega\right\}_{\mathrm{connected},n}\, ,
\label{Eq:CCeqn}
\end{eqnarray}
where the valence correlation energy
\begin{equation}
\delta E_v= \langle\Psi_v^{(0)}|G \Omega|\Psi_v^{(0)}\rangle \label{Eq:CorrEnergy} \, ,
\end{equation}
and $Q=1-|\Psi_v^{(0)}\rangle \langle\Psi_v^{(0)}|$ is a projection
operator. Notice that only connected diagrams are retained on the
r.h.s of the equation, r.h.s. diagrams being of the the same
topological structure as clusters $\left(K\right)_n$. The resulting
CC equations for the core clusters do not depend on the valence
state.


Although the CC approach is strictly exact, in practical
applications the full cluster operator $K$ is truncated at a certain
level of excitations, e.g., at single and double excitations (CCSD
method). In particular, for univalent atoms the CCSD parametrization
may be represented as
\begin{equation}
 K^\mathrm{SD}= S_c + D_c + S_v + D_v =
 \sum_{ma} \rho_{ma}    \,  a^\dagger_m a_a +
 \frac{1}{2!}  \sum_{mnab} \rho_{mnab} \, a^\dagger_m a^\dagger_n a_b a_a +
  \sum_{m \ne v} \rho_{mv}   \,   a^\dagger_m a_v +
 \sum_{mna} \rho_{mnva}    \,  a^\dagger_m a^\dagger_n a_a a_v \, ,
\end{equation}
The cluster  amplitudes $\rho_{\cdots}$ are to be determined from
the Eq.(\ref{Eq:CCeqn}).

A {\em linearized} version of the CCSD method discards non-linear
terms in the expansion of exponent in Eq.~(\ref{Eq:CCOmega}) of the
coupled-cluster parametrization, i.e.,  $\Omega^\mathrm{SD} \equiv 1
+ K^\mathrm{SD}$. This leads to discarding disconnected excitations
from the exact many-body wave function. We will refer to this
approximation simply as singles-doubles (SD) method. For
alkali-metal atoms the SD method was employed previously by the
Notre Dame
group~\cite{BluJohLiu89,BluJohSap91,SafDerJoh98,SafJohDer99}. The
resulting SD equations are written out in Ref.\cite{BluJohLiu89}. A
typical {\em ab initio} accuracy attained for properties of heavy
alkali-metal atoms is at the level of 1\%.

Successive iterations of the CC equations (\ref{Eq:CCeqn}) recover
the traditional order-by-order MBPT. As discussed in
Ref.\cite{BluJohLiu89}, the core and valence doubles appear already
in the first order in the residual interaction $G$:
\begin{eqnarray}
 \rho_{mnab} &\approx& \frac{ g_{mnab}}
{\varepsilon_a + \varepsilon_b-\varepsilon_m-\varepsilon_n} \, , \\
\rho_{mnva} &\approx& \frac{ g_{mnva}}
{\varepsilon_v + \varepsilon_a-\varepsilon_m-\varepsilon_n} \, . \label{Eq:DoublesLowOrder}
\end{eqnarray}
Valence and core singles appear at the second iteration of the CC
equations and are effectively of the second order in $G$. We will
employ this ``effective order'' classification to develop our
approximation to the CC equations.

\subsection{Triple excitations. Motivating discussion}

Certainly the truncation of the CC expansion leads to a neglect of
many-body diagrams containing excitations beyond singles and
doubles. For example, both the SD and the CCSD methods recover all
the diagrams for valence energies through the second order of MBPT,
but start missing diagrams associated with valence triple
excitations in the third order~\cite{BluJohLiu89}. Similarly, for
contributions to matrix element of a one-body (e.g., electric dipole
) operator, the SD method subsumes all the diagrams through the
third order but misses approximately half of the diagrams in the
fourth order of MBPT. The omitted fourth-order diagrams are entirely
due to triple and disconnected quadruple
excitations~\cite{DerEmm02}. Our group has carried out calculations
of these 1,648 complementary diagrams for Na~\cite{CanDer04} and
Cs~\cite{DerPor05}. Close examination of our computed complementary
diagrams reveals a high ( a factor of a hundred ) degree of
cancelation between different contributions. Such cancelations could
lead to a poor convergence of the MBPT series. Poor convergence
calls for an all-order summation scheme and this is what we address
here. The resulting formalism will recover the dominant fourth-order
contributions to matrix elements and all third-order MBPT
contributions to the valence energies in a nonperturbative fashion.

\begin{figure}[h]
\begin{center}
\includegraphics*[scale=0.6]{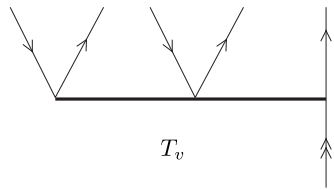}
\caption{Diagrammatic representation of valence triple excitations.
Double-headed arrow represents valence state.
\label{Fig:Tv}}
\end{center}
\end{figure}

The next systematic step in improving the SD
method would be an additional inclusion of triple excitations
\begin{eqnarray}
T_c &=&
  \frac{1}{12}  \sum_{mnrabc} \rho_{mnrabc} \, a^\dagger_m a^\dagger_n
  a^\dagger_r a_c a_b a_a  \, , \\
T_v &=&  \frac{1}{6}  \sum_{mnrab} \rho_{mnrvab} \, a^\dagger_m
a^\dagger_n
  a^\dagger_r  a_b a_a a_v
\end{eqnarray}
into the cluster operator $K$ (see Fig.~\ref{Fig:Tv}). However,
considering the present state of available computational power, the
full incorporation of triples (specifically, core triples) seems to
be yet not practical for heavy atoms.

To motivate more accurate, yet practical extension of the SD method,
we consider numerical results for the reduced electric-dipole matrix
elements of $3s_{1/2}-3p_{1/2}$ transition in Na~\cite{CanDer04}.
From Table~I of that paper, we observe that the contributions from
{\em valence} triples $T_v$ (total $-4.4 \times 10^{-3}$) and
non-linear doubles (disconnected quadruples) $D_{nl}$ (total  $1.3
\times 10^{-3}  $) are much larger than those from {\em core}
triples $T_c$ (total $8 \times 10^{-5}$). Similar conclusion can be
drawn from our  calculations for heavier Cs atom~\cite{DerPor05}.
Because of this observation  we will discard core triples and
incorporate the {\em valence triples} into the SD formalism. We will
refer to this method as SDvT approximation. Contributions of core
triples to matrix elements are treated in this work perturbatively.

In addition to triples, we will  include effects from disconnected
excitations. The relevant diagrams contribute at the same level as
the valence triples and the full treatment of disconnected
excitations will recover a part of the otherwise missing sequence of
random-phase-approximation diagrams (see also discussion in
Ref.~\cite{DerPor05}). The resulting approximation will be referred
to as CCSDvT method.

\section{Formalism}
\label{Sec:Formalism}
Below we write down the CC equations for
cluster amplitudes $\rho$ in the CCSDvT approximation. The equations
in the SD approximation are presented in Ref.~\cite{BluJohLiu89}. We
retain convention for the single and doubles from that paper and
focus on additional terms due to valence triples and disconnected
excitations. Some of the equations involving triple excitations were
given in Ref.~\cite{SafDerJoh98,SafJohDer99}; we use a different
convention for the triples amplitudes.

\subsection{Valence triples}
In the following, we employ fully antisymmetrized valence triples
amplitude $\tilde{\rho}_{mnrvab}$. The object
$\tilde{\rho}_{mnrvab}$ is antisymmetric with respect to any
permutation of the indices $mnr$ or $ab$, e.g.,
\begin{equation}
\tilde{\rho}_{mnrvab}=-\tilde{\rho}_{nmrvab}=-\tilde{\rho}_{mnrvba} =
\tilde{\rho}_{mrnvba} =\ldots \, . \label{Eq:TvAntiSym}
\end{equation}
It is straightforward to demonstrate that the contribution to the wave
operator (and therefore all the resulting equations)
can be expressed in terms of this antisymmetrized object. Explicitly,
\begin{equation}
T_v =  \frac{1}{12}  \sum_{mnrab} \tilde{\rho}_{mnrvab} \, a^\dagger_m a^\dagger_n
  a^\dagger_r  a_b a_a a_v \, .
\end{equation}
Computationally the use of $\tilde{\rho}_{mnrvab}$ substantially
reduces  storage requirements, as it is sufficient to store ordered
amplitudes with $m > n > r$ and $a > b$ only. In the equations
below, we will also  use antisymmetrized combinations for doubles
$\tilde{\rho}_{mnab} = \rho_{mnab} - \rho_{mnba} = \rho_{mnab} -
\rho_{nmab}$, $\tilde{\rho}_{mnva} = \rho_{mnva} - \rho_{nmva}$, and
for the Coulomb matrix elements $\tilde{g}_{ijkl} = g_{ijkl} -
g_{ijlk}$.

\begin{figure}[h]
\begin{center}
\includegraphics*[scale=0.6]{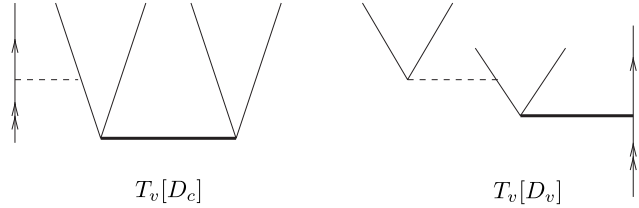}
\caption{ Representative contributions to the r.h.s of the valence
triples equation. Horizontal dashed line denotes Coulomb interaction and solid
lines --- cluster amplitudes.
 \label{Fig:TvDcDv}}
 \end{center}
\end{figure}

From the general Eq.(\ref{Eq:CCeqn}) we obtain symbolically
\begin{eqnarray}
\lefteqn{ \left(
\varepsilon_{a}+\varepsilon_{b}+\varepsilon_{v}-\varepsilon
_{m}-\varepsilon_{n}-\varepsilon_{r}+\delta E_{v}\right)
\tilde{\rho}_{mnrvab} = }\nonumber\\
& & T_v[D_c]+ T_v[D_v]  + T_v[T_v] + T_v[T_c] + \mathrm{nonlinear} \,.
\end{eqnarray}
Here contribution $T_v[D_c]$ denotes effect of core doubles on
valence triples, the remaining terms defined in a similar fashion.
In this work we include only contributions $T_v[D_c]$ and $T_v[D_v]$
(see Fig.~\ref{Fig:TvDcDv}) and omit the effect of valence and core
triples on valence triples ( $T_v[T_v]$ and $T_v[T_c]$) and
nonlinear CC contributions. Compared to the $T_v[D_v]$ and
$T_v[D_c]$ contributions, these are higher order (and
computationally expensive) effects. Explicitly,
\begin{eqnarray}
T_v[D_c] & = & - \underset{c}{\sum}\left(
\widetilde{g}_{mcva}\widetilde{\rho}_{nrcb}-\widetilde{g}_{mcvb}%
\widetilde{\rho}_{nrca}+\widetilde{g}_{ncva}\widetilde{\rho}_{rmcb}%
-\widetilde{g}_{ncvb}\widetilde{\rho}_{rmca}+\widetilde{g}_{rcva}%
\widetilde{\rho}_{mncb}-\widetilde{g}_{rcvb}\widetilde{\rho}_{mnca}
\right)  \nonumber \\
&+&  \underset{s}{\sum}\left(
\widetilde{g}_{nrsv}\widetilde{\rho}_{msab}+\widetilde{g}_{rmsv}%
\widetilde{\rho}_{nsab}+\widetilde{g}_{mnsv}\widetilde{\rho}_{rsab}%
\right) \, , \\
T_v[D_v] & = &   \underset{c}{\sum} \left(
\widetilde{g}_{mcab}\widetilde{\rho}_{nrvc}+\widetilde{g}_{ncab}%
\widetilde{\rho}_{rmvc}+\widetilde{g}_{rcab}\widetilde{\rho}_{mnvc} \right) \nonumber \\
& + &\underset{s}{\sum}\left(
\widetilde{g}_{nrsb}\widetilde{\rho}_{msva}-\widetilde{g}_{nrsa}%
\widetilde{\rho}_{msvb}+\widetilde{g}_{rmsb}\widetilde{\rho}_{nsva}%
-\widetilde{g}_{rmsa}\widetilde{\rho}_{nsvb}+\widetilde{g}_{mnsb}%
\widetilde{\rho}_{rsva}-\widetilde{g}_{mnsa}\widetilde{\rho}_{rsvb} \right) \,.
\label{Eq:rho_mnrvab}
\end{eqnarray}
 Notice that the matching of diagrams in
Eq.(\ref{Eq:CCeqn})  is generally not unique; we require that the
r.h.s. of the above equation is fully antisymmetrized as the
amplitude $\tilde{\rho}_{mnrvab}$ on the l.h.s.; such procedure is
unique and corresponds to a projecton of the CC equations onto the
many-body state $a^\dagger_m a^\dagger_n a^\dagger_r  a_b a_a
|0_c\rangle$. Also from these equations we immediately observe that
the triples enter the many-body wave function in the effective
second order of MBPT, as the doubles enter in the first order in
$G$, Eq.(\ref{Eq:DoublesLowOrder}).

\subsection{Modifications to SD equations and valence energies}

Here we present CC equations for correlation energy $\delta E_v$,
valence singles $\rho_{mv}$, and for valence double $\rho_{mnva}$
cluster amplitudes. In formulas below we write $\mathrm{SD}$ to
denote contributions in the singles-doubles approximations tabulated
in Refs.~\cite{BluJohLiu89,SafDerJoh98}. As to the core amplitudes,
they will be determined in the SD approximation (i.e. we do not
include non-linear CC terms and core triples).

\begin{figure}[h]
\begin{center}
\includegraphics*[scale=0.5]{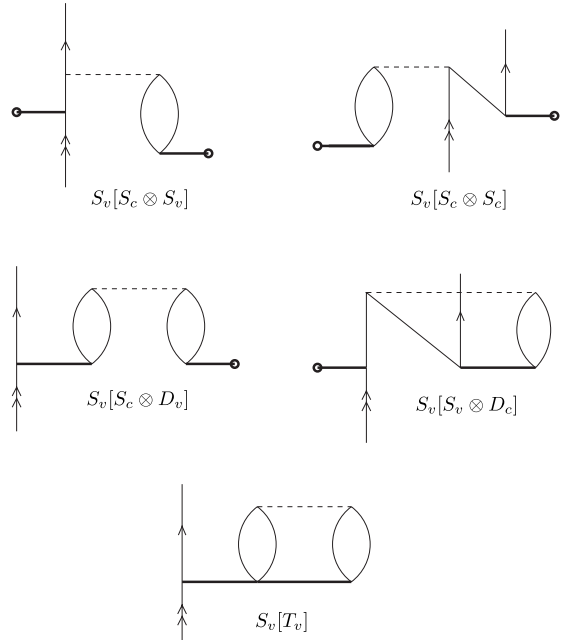}
\caption{Sample contributions of triples and
disconnected excitations to the valence singles equation.
 \label{Fig:Sv}}
 \end{center}
\end{figure}

Topological structure of the valence singles equation is
\begin{eqnarray}
\lefteqn{\left( \varepsilon_{v}-\varepsilon_{m}+\delta E_v \right)
\rho_{mv}  = \mathrm{SD} + } \nonumber \\
& &
S_v[S_c \otimes S_v] + S_v[S_c \otimes S_c] +
S_v[S_c \otimes D_v]  +  S_v[S_v \otimes D_c] +
S_v[T_v] \, ,
\label{Eq:rho_mvTopo}
\end{eqnarray}
where notation $(K)_n\left[ (K)_p \otimes (K)_m\right]$ stands for a
contribution from a disconnected $(p+m)$--fold excitation (resulting
from a product of clusters $(K)_p$ and $(K)_m$) to the cluster
$(K)_n$. We do not include cubic non-linear term $S_v[S_c \otimes
S_c \otimes S_v]$. Explicitly,
\begin{eqnarray}
S_v[S_c \otimes S_v] &=& \underset{anr}{\sum} \tilde{g}_{amnr} \rho_{na} \rho_{rv} \, , \\
S_v[S_c \otimes S_c] &=& \underset{abn}{\sum} \tilde{g}_{abnv} \rho_{ma} \rho_{nb} \, , \\
S_v[S_c \otimes D_v] &=& \underset{abnr}{\sum} \tilde{g}_{abnr} \left(
 \rho_{mb} \rho_{nrva} - \rho_{nb} \tilde{\rho}_{mrva} \right) \, ,\\
S_v[S_v \otimes D_c] &=& - \underset{abnr}{\sum} \tilde{g}_{abnr} \rho_{nv} \rho_{mrab} \, , \\
S_v[T_v] &=& \frac{1}{2}\underset{abnr}{\sum}g_{abnr}\tilde{\rho}_{mnrvab} \,.
\label{Eq:rho_mv}
\end{eqnarray}
Representative diagrams are shown in Fig.~\ref{Fig:Sv}.

\begin{figure}[h]
\begin{center}
\includegraphics*[scale=0.6]{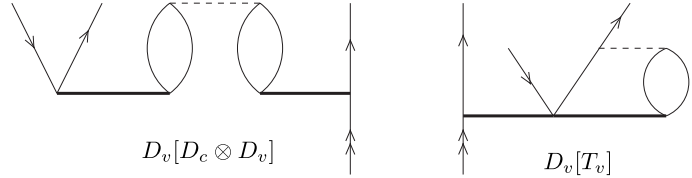}
\caption{ Effects of
disconnected and valence triple excitations on valence doubles.
 \label{Fig:Dv}}
 \end{center}
\end{figure}

Valence doubles equation for $\rho_{mnva}$
can be symbolically represented as (see Fig.~\ref{Fig:Dv})
\begin{eqnarray}
\lefteqn{ \left(\varepsilon_{v}+\varepsilon_{a}-
\varepsilon_{m}-\varepsilon_{n}+\delta E_{v}\right)
\rho_{mnva}  = \mathrm{SD} +}\nonumber \\
&& D_v[S_c \otimes S_v] + D_v[S_c \otimes S_c] +  \nonumber\\
&& D_v[S_c \otimes D_v]  +  D_v[S_v \otimes D_c] + D_v[S_c \otimes D_c] + \label{Eq:DvSymb}\\
&& D_v[D_c \otimes D_v] +   D_v[S_c \otimes T_v] + D_v[S_v \otimes T_c] +
 D_v[T_v] \, . \nonumber
\end{eqnarray}
Contribution $D_v[D_c \otimes D_c]$ is topologically impossible and we omit cubic and
higher-degree nonlinear terms like $D_v[S_c \otimes S_c \otimes S_v]$,
$D_v[S_c \otimes S_c \otimes D_v]$, and $D_v[S_v \otimes S_c \otimes S_c \otimes S_c ]$.

Explicitly,
\begin{align*}
D_v[D_c \otimes D_v]
&= \underset{bcrs}{\sum} g_{bcrs} \left\{
   \rho_{rsva} \rho_{mnbc}  +
   \frac{1}{2}\tilde{\rho}_{msva} \rho_{nrbc}  \right. \nonumber \\
& \left.
    + \frac{1}{2}\tilde{\rho}_{snva} \rho_{mrbc} +
    \tilde\rho_{rsvb} \rho_{nmac} + \tilde\rho_{rsab} \rho_{mnvc} \right\}
    \nonumber \, \\
& -\underset{bcrs}{\sum} \tilde g_{bcrs} \tilde{\rho}_{msvb} \tilde{\rho}_{nrac}, \\
D_{v}[S_{v}\otimes D_{c}]& =-\sum_{brs}\tilde{g}_{bmrs}\,\rho_{rv}%
\,\tilde{\rho}_{nsab}+\sum_{bcr}g_{bcar}\,\rho_{rv}\,\rho_{nmbc} \, , \\
D_{v}[S_{c}\otimes D_{v}]  &  = \frac{1}{2}\sum_{brs}\tilde{g}_{bnrs}%
\,\rho_{rb}\,\tilde{\rho}_{msva}\,\,\,-\frac{1}{2}\sum_{brs}\tilde{g}%
_{bmrs}\,\rho_{rb}\,\tilde{\rho}_{nsva}\,\,\\
&  +\frac{1}{2}\sum_{brs}g_{bmrs}\,\rho_{nb}\,\tilde{\rho}_{rsva}\,\,-\frac
{1}{2}\sum_{brs}g_{bnrs}\,\rho_{mb}\,\tilde{\rho}_{rsva}\\
&  -\sum_{brs}\tilde{g}_{bnrs}\,\rho_{ra}\,\tilde{\rho}_{msvb}-\sum
_{bcr}\tilde{g}_{bcar}\,\rho_{rc}\,\rho_{mnvb}-\sum_{bcr}\tilde{g}%
_{bcar}\,\rho_{nc}\,\tilde{\rho}_{rmvb}\,  ,\\
D_{v}[S_{c}\otimes D_{c}]&=-\sum_{bcr}\tilde{g}_{bcvr}\,\rho_{rc}\,\rho
_{nmab}\,-\sum_{bcr}\tilde{g}_{bcvr}\,\rho_{mc}\,\tilde{\rho}_{rnab}%
\,+\,\sum_{bcr}g_{bcvr}\,\rho_{ra}\,\rho_{mnbc} \, , \\
D_{v}[S_{c}\otimes S_{v}]  &  =\sum_{br}\tilde{g}_{bnar}\,\rho_{mb}\,\rho
_{rv}\,\,+\sum_{rs}g_{mnrs}\,\rho_{rv}\,\rho_{sa}\, , \\
D_{v}[S_{c}\otimes S_{c}]  &  =\sum_{br}\tilde{g}_{bmvr}\,\rho_{nb}\,\rho
_{ra}+\sum_{bc}g_{bcav}\,\rho_{mc}\,\rho_{nb}\, .
\end{align*}

The effect of valence triples on valence doubles reads
\begin{eqnarray*}
D_v[T_v] &=& -\frac{1}{2} \underset{rbc} {\sum} \left(
    g_{bcar} \tilde{\rho}_{mnrvbc}+ g_{bcvr} \tilde{\rho}_{nmrabc} \right)
  \\
   &+& \frac{1}{2}\underset{rsb}{\sum} \left(
    g_{bnrs} \tilde{\rho}_{msrvab}+
    g_{bmrs} \tilde{\rho}_{snrvab} \right) \,.
\end{eqnarray*}

Finally, the valence correlation energy may be represent as
\begin{equation}
\delta E_v = \delta E_\mathrm{SD} + \delta E_\mathrm{CC} + \delta E_\mathrm{vT} \,,
\label{Eq:delta_Ev}
\end{equation}
with
\begin{eqnarray}
\delta E_\mathrm{CC} &=&
\underset{anr}{\sum}\ \tilde{g}_{avnr} \rho_{na} \rho_{rv} +
\underset{abn}{\sum}\ \tilde{g}_{abnv} \rho_{va} \rho _{nb} \label{Eq:dECC} \\
& + & \underset{abnr}{\sum}\ \tilde{g}_{abnr}
\left[ \rho_{vb} \rho_{nrva} - \rho_{nb} \tilde{\rho}_{vrva} -
       \rho_{nv} \rho_{vrab} \right] \, ,                              \nonumber  \\
\delta E_\mathrm{vT} &=& \frac{1}{2}\underset{abmn}{\sum}\ g_{abmn}
\tilde{\rho}_{vmnvab}\,. \label{Eq:dEvT}
\end{eqnarray}
Topological structure of  contributions to energy is similar to the
terms on the r.h.s of the valence singles
equation~(\ref{Eq:rho_mvTopo}). Here correction $\delta
E_\mathrm{CC}$ comes from non-linear CC contributions and $\delta
E_\mathrm{vT}$ is due to valence triples.

\subsection{Normalization}
The CC wave function is derived using the intermediate
normalization, $\langle \Psi_v^{(0)} | \Psi_v \rangle=1$ and in
calculating the atomic properties based on the CC wave function, one
needs to renormalize it. In calculations of matrix elements one
requires the valence part of the normalization, $N_v = \langle
\Psi_v | \Psi_v \rangle_\mathrm{val,connected}$. We obtain
\begin{eqnarray}
N_v =
 \mathrm{SD}+\underset{mnab} {\sum} \rho_{mnab} \,%
\tilde{\rho}_{vmnvab}+\frac{1}{12}\underset{mnrab}{\sum}(\tilde{\rho}%
_{mnrvab})^{2} \,.
\label{Eq:N}
\end{eqnarray}
The last term in the equation above is quadratic in valence triples
(i.e., it is of the fourth effective order) and we  will neglect it
in the following.

\subsection{Matrix elements of one-body operator}
\label{Sec:MelFormalism}
Finally, we consider  matrix elements of a one-body operator
$Z = \sum_{ij} z_{ij} a_i^\dagger a_j$ between two
CC states $|\Psi_v \rangle$ and $|\Psi_w \rangle$.
Taking into account renormalization, this matrix element
can be defined as
\begin{equation}
\mathcal{M}_{wv} \equiv \frac{ \langle \Psi_w |Z| \Psi_v \rangle
}{\sqrt{N_w N_v}}
\end{equation}
As it was shown in Ref.~\cite{BluJohLiu89} all disconnected
diagrams in the numerator and denominator of this expression cancel, leading
to
\begin{eqnarray}
\mathcal{M}_{wv} = \frac{\left(  Z_{wv}^\mathrm{val}
\right)_{\mathrm{conn}} }{\left\{  \left[  1+\left(
N_{v}^\mathrm{val} \right)_{\mathrm{conn} }\right]  \left[  1+\left(
N_{w}^\mathrm{val} \right)_{\mathrm{conn} }\right]  \right\} ^{1/2}}
\, .
\label{Eq:Zconn}
\end{eqnarray}
We discarded valence-independent contribution, as it vanishes for non-scalar
operators. To unclutter the notation below we simply write
\begin{eqnarray}
 Z_{wv} &\equiv& (Z_{wv}^\mathrm{val})_{\mathrm{conn}}, \nonumber \\
 N_v &\equiv& (N_v^\mathrm{val})_{\mathrm{conn}} \,.
\end{eqnarray}

\citet{BluJohLiu89} tabulated 21 contributions to the matrix elements  in the SD
approximation.
These SD corrections are mainly due to (i)
random-phase-approximation (RPA) diagram proportional to a product
of $Z$ and $D_v$ and (ii) the Brueckner-type (core-polarization)
diagram proportional to the product of $Z$ and $S_v$. In
Ref.~\cite{DerPor05} we additionally included modifications to
$\mathcal{M}_{wv}$ caused by non-linear terms in the CC wave
function. We have devised a re-summation scheme that is equivalent
to ``dressing'' of lines and vertices of the SD diagrams (see also
Ref.~\cite{MarYnn90}).

Including valence triples leads to additional direct contributions,
 $Z_{wv}={\rm SD} + Z_{wv}^{\left(T_v \right)}$. We obtain
\begin{eqnarray}
Z_{wv}^{\left(T_v \right) } &=&  \sum_{k=1}^{7} Z_{wv}^{\left(T_v,k \right) }\, , \label{Eq:ZT}\\
Z_{wv}^{\left(T_v,1 \right) } &=&
\underset{abmn}{\sum}\rho_{ma}^*\tilde{\rho}_{wmnvab}  z_{bn} + {\rm h.c.s.},  \\
Z_{wv}^{\left( T_v,2\right) } &=&
-\frac{1}{2}\underset{abcmn}{\sum} \tilde{\rho}_{mnba}^* \tilde{\rho}_{wnmvcb}
z_{ca} + {\rm h.c.s.}, \\
Z_{wv}^{\left( T_v,3\right) } &=&
\frac{1}{4}\underset{abmnr}{\sum } \tilde{\rho}_{mnab}^* \tilde{\rho}_{rmnvab}
z_{wr} + {\rm h.c.s.},  \\
Z_{wv}^{\left( T_v,4\right) } &=&
\frac{1}{2}\underset{abmnr}{\sum }\tilde{\rho}_{mnab}^* \tilde{\rho}_{wrmvba}
z_{nr} + {\rm h.c.s.},  \\
Z_{wv}^{\left( T_v,5\right) } &=&
-\frac{1}{2}\underset{abmnr}{\sum } \tilde{\rho}_{mnwb}^* \tilde{\rho}_{rmnvab}
z_{ar} + {\rm h.c.s.},  \\
Z_{wv}^{\left( T_v,6\right) } &=&-\frac{1}{6}\underset{abcmnr}{\sum }\tilde{\rho}%
_{mnrwcb}^{\ast }\tilde{\rho}_{mnrvab} z_{ac},  \\
Z_{wv}^{\left( T_v,7\right) } &=&\frac{1}{4}\underset{abmnrs}{\sum }\tilde{\rho}%
_{mnrwab}^{\ast }\tilde{\rho}_{snrvab} z_{ms} \,.%
\end{eqnarray}
In these expressions, abbreviation h.c.s. stands for a Hermitian
conjugation of the preceding term with a simultaneous swap of the
valence indices $w \leftrightarrow v$. As discussed in
Ref.~\cite{DerEmm02}, valence triples start contributing in the
fourth order of MBPT for matrix elements; these contributions
correspond to terms $Z_{wv}^{\left( Tv,k\right) }, k=2-5$. We
presently discard terms \#6 and \#7 that are quadratic in triple
excitations.

\subsection{Symmetries and reduced triples}
Relativistic one-particle orbitals $i$ are characterized by the
principle quantum number $n_i$, the total angular momentum $j_i$,
it's projection $m_i$ and the orbital angular momentum $l_i$. The
summations over magnetic quantum numbers are carried out
analytically, substantially reducing the number of coefficients. A
dependence of valence triples on magnetic quantum numbers may be
parameterized as (we use angular diagrams, see, e.g.,
Ref.~\cite{LinMor86})
\begin{equation}
\tilde \rho_{mnrvab}=\sum_{LL^{\prime}h}%
\raisebox{-6ex}{\includegraphics[
scale=0.6
]{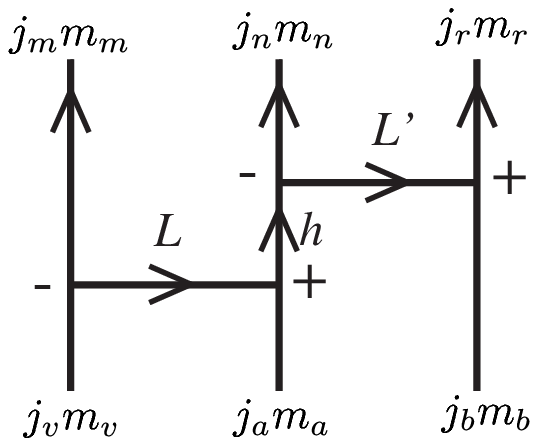}}
\tilde{F}_{LL^{\prime}h}\left(  mnr \, vab\right) \,,
\label{Eq:vTAngular}
\end{equation}
where $h$ is a half-integer coupling angular momentum and $L$ and
$L^{\prime}$ are integer coupling momenta. The ``reduced triples''
$\tilde{F}_{LL^{\prime}~h}\left(  mnrvab\right)$ do not depend on magnetic
quantum numbers.

Selection rules for various angular momenta characterizing
reduced triples follow from properties
of the $3j$-symbols in the angular diagram~(\ref{Eq:vTAngular}). In addition,
the atomic Hamiltonian is invariant under parity transformation, leading to
an additional parity selection rule  $l_m+l_n+l_r+l_v+l_a+l_b = \mathrm{even \, integer}$
for a triple amplitude $\tilde \rho_{mnrvab}$.

Owing to the antisymmetric properties of the triples,
Eq.(\ref{Eq:TvAntiSym}), it is sufficient to store reduced triples
with $(n_m \varkappa_m) \ge (n_n \varkappa_n)\ge (n_r \varkappa_r)$
and $(n_a \varkappa_a) \ge (n_b \varkappa_b)$, where $\varkappa =
(l-j)(2j+1)$. The reduced triples with other combinations of
arguments can be related to the ordered set via symmetry properties.
For example,
\begin{equation}
\tilde{F}_{LL^{\prime}h}\left(  mnr \, vba\right)  =(2h+1)\,(2L^{\prime}+1)
\,\sum_{h'K}\left\{
\begin{array}
[c]{ccc}%
j_b & h & L\\
j_r & L^{\prime} & j_a\\
K & j_n & h'
\end{array}
\right\}  \,{\left(  -1\right)  }^{h+h'+K+L^{\prime}}\,{\tilde{F}%
}_{L K h'}(mnr \, vab)\,\,.
\end{equation}
There are 11 such index-swapping relations for reduced valence triples.

\section{Numerical results and discussion}
\label{Sec:Results}
To reiterate discussion so far, we derived
algebraic expressions in the CCSDvT formalism, which includes
valence triples and a subset of disconnected excitations. We also
carried out angular reduction of these expressions and developed a
numerical code. In this section we present our {\em ab initio}
results for properties of $3s$, $3p_{1/2}$, and $3p_{3/2}$ states of
atomic sodium. Results for removal energies are presented in
Section~\ref{Sec:Energies} and for dipole matrix elements and HFS
constants $A$ in Section~\ref{Sec:Mels}.

Before presenting the results, let us briefly describe our numerical
code. It is an extension of the relativistic SD
code~\cite{SafDerJoh98} which employs  B-spline basis set. This
basis  numerically approximates complete set  of single-particle
atomic states. Here we use 35 out of 40 positive-energy
($\varepsilon_i > -m_e c^2$) basis functions. Basis functions with
$l_{\rm max} \le 6$ are used for singles and doubles. For triples we
employ a more limited set of basis functions with $l_{\rm max}(T_v)
\le 4$. Excitations from all core sub-shells are included in the
calculations. Numerically we found that this choice is a reasonable
trade-off between storage and overall numerical accuracy (after all,
triples affect computed properties at $\sim 1$\% level.) The results
presented in this Section will include basis set extrapolation
correction, which is obtained by computing SD properties with
increasingly larger basis sets and interpolating them to $l=\infty$.
The CC equations were solved iteratively. We notice that the
reported calculations can be carried out in the memory of a modern
high-end personal workstation: storing reduced valence triples in a
single precision required about 900 Mb for $s_{1/2}$ states and 1.5
Gb for $p_{3/2}$ states (the latter involve more angular channels).

\subsection{Energies}
\label{Sec:Energies}
Computed removal energies of
$3s$, $3p_{1/2}$, and $3p_{3/2}$ states of atomic sodium are presented in
Table~\ref{Tab:Na_E}. The dominant contribution to the energies comes
from the DHF values. The remaining (correlation) contribution
is given by Eq.~(\ref{Eq:delta_Ev}). We computed this correlation correction
in several approximations: SD, SDvT, CCSD, and, finally, CCSDvT.

\begin{table}[h]
\caption{Contributions to removal energies of $3s$, $3p_{1/2}$, and
$3p_{3/2}$ states for Na in cm$^{-1}$ in various approximations. A
comparison with previous CC-type calculations and experimental
values is presented in the lower panel. }
\label{Tab:Na_E}
\begin{ruledtabular}
\begin{tabular}{lddd}
& \multicolumn{1}{c}{$3s$} & \multicolumn{1}{c}{$3p_{1/2}$} & \multicolumn{1}{c}{$3p_{3/2}$}\\
\hline
 $E_{\rm DHF}$              & 39951.6 & 24030.4 & 24014.1 \\[2ex]
\multicolumn{4}{c}{SD} \\
 $\delta E_{\rm SD}$        &  1488.8 &   463.9 &  460.6 \\
 $E^{\rm tot}_{\rm SD}$     & 41440.3 & 24494.3 & 24474.7 \\[2ex]
\multicolumn{4}{c}{SDvT} \\
 $\delta E_{\rm SD}^{\rm indir}$ & 79.7  &  28.9  &    28.4 \\

 $\delta E_{\rm vT}$         &    25.4 &    4.8 &      4.7 \\
 $E^{\rm tot}_{\rm SDvT}$    &  41545.5 &  24528.0 &  24507.8  \\[2ex]
\multicolumn{4}{c}{CCSD} \\
 $\delta E_{\rm SD}^{\rm indir}$ &-57.0 &  -20.0 &     -18.4 \\ 
 $\delta E_{\rm CC}$         &   -17.5   &   -7.4  &    -7.4  \\
 $E^{\rm tot}_{\rm CCSD}$    & 41365.9   & 24466.9 & 24448.9  \\[2ex]
\multicolumn{4}{c}{CCSDvT} \\
$\delta E_{\rm SD}^{\rm indir}$ & 16.8  &   6.8  &    7.9 \\
$\delta E_{\rm vT}$        &     23.7 &      4.5 &      4.4 \\
$\delta E_{\rm  CC}$        &    -18.4 &     -8.0 &     -8.0 \\
$E^{\rm tot}_{\rm CCSDvT}$
                            &  41462.5 &  24497.6 &  24479.1  \\ \hline
\multicolumn{4}{c}                    {Other works}      \\
SD(pvT) ~\cite{SafJohDer99}
                            &  41447.3 &  24493.9 &  24476.7  \\
CCSD~\cite{EliKalIsh94}
                            &  41352   &  24465   &         \\
{\rm $E_{\rm experim}$}~\footnotemark[1]
                            &  41449.6 &  24493.4 &  24476.2  \\
\end{tabular}
\end{ruledtabular}
\footnotemark[1]{These values are from spectroscopic data
compiled by NIST~\protect\cite{NIST_ASD}}
\end{table}

First we list correlation energies $\delta E_{\rm SD}$ obtained in
the SD approximation. The results contain basis set extrapolation
corrections from Ref.~\cite{Saf00}. The extrapolation corrections
increase the removal energies by  5.1 cm$^{-1}$ for the $3s$ state,
1.9 cm$^{-1}$ for the $3p_{1/2}$ state, and 0.8 cm$^{-1}$ for the
$3p_{3/2}$. Total removal energy is $E^\mathrm{tot}_\mathrm{SD} =
E_\mathrm{DHF} + \delta E_\mathrm{SD}$. At the next step (SDvT) we
include valence triple excitations, i.e., in the CC equations in
addition to the SD terms we incorporate terms with amplitudes
$\tilde{\rho}_{mnrvab}$. It is instructive to distinguish direct and
indirect  $\delta E_{\rm SD}^{\rm indir}$ effects of these
excitations. The direct effect of triples is $\delta E_\mathrm{vT}$,
Eq.(\ref{Eq:dEvT}), while indirect effect is a modification of
$\delta E_\mathrm{SD}$ due to  effect of triples through coupling to
singles and doubles.  In this case, the indirect contribution is
defined as $\delta E_{\rm SD}^{\rm indir} = \delta E_{\rm SD}[{\rm
SDvT}] -  \delta E_{\rm SD}[{\rm SD}]$. We list the two types of
contributions in the Table and it is clear that for all the states
both contributions add constructively, and for all the considered
approximations the indirect contribution dominates over the direct
one. The total removal energy in the SDvT approximation  is
$E^\mathrm{tot}_\mathrm{SDvT} = E_\mathrm{DHF} + \delta
E_\mathrm{SD} + \delta E_{\rm SD}^{\rm indir} + \delta E_\mathrm{vT}
$. The totals for other approximations are defined in a similar way.

As we move to the CCSD approximation in Table~\ref{Tab:Na_E}, we
notice that here the corrective terms $\delta E_{\rm SD}^{\rm
indir}$ and $\delta E_{\rm CC}$ decrease the removal energies, while
for the SDvT case the corrections increased $E^\mathrm{tot}$. In
both cases the resulting total energies $E^\mathrm{tot}$ were moved
away from the experimental values. Since the effects of disconnected
and triple excitations are comparable and opposite in sign,
simultaneous treatment of the two effects is required. The results
of such treatment are listed under CCSDvT heading in the Table.
Compared to the CCSD and SDvT approximations, the CCSDvT results
move into a closer, 0.01-0.03\%, agreement with the experimental
values.

Comparison with the previous CC-type calculations of Na removal
energies is presented in the lower panel of the
Table~\ref{Tab:Na_E}. SD(pvT) approximation denotes results obtained
with a scheme originally proposed in Ref.~\cite{BluJohSap91}. In
this scheme: (i) starting from the SDvT approximation, one keeps vT
contributions in the equation for valence singles and valence
energies (i.e., $D_v[T_v]$ effect is neglected) ; (ii) triples are
approximated by
\[
\tilde{\rho}_{mnrvab} \approx \frac{T_v[D_c]+ T_v[D_v] }
{\varepsilon_{a}+\varepsilon_{b}+\varepsilon_{v}-\varepsilon
_{m}-\varepsilon_{n}-\varepsilon_{r}+\delta E_{v}} ;
\]
(iii) to avoid expensive storing of valence triples, in the
$\rho_{mv}$ equation the triples denominators
$\left({\varepsilon_{a}+\varepsilon_{b}+\varepsilon_{v}-\varepsilon
_{m}-\varepsilon_{n}-\varepsilon_{r}+\delta E_{v}}\right) $ are
replaced by an approximate combination
$\left({\varepsilon_{a}+\varepsilon_{b}-\varepsilon_{n}-\varepsilon_{r}}\right)$.
In  this approximation $S_v[T_v]$ effect is effectively
overemphasized ( for the ground state $\varepsilon_{v} <
\varepsilon_{m}$ ). In the expression for the energy, $\delta
E_\mathrm{vT}$, Eq.(\ref{Eq:dEvT}), triples enter as
$\tilde{\rho}_{vmnvab}$ and the above replacement of denominators is
more algebraically justified. Nevertheless, we found a substantial
(a factor of three) disagreement between $\delta E_\mathrm{vT}$
corrections obtained in our (more complete) SDvT and SD(pvT)
approximations.

To understand the origin of this large disagreement, we have
compared individual contributions to $\delta E_\mathrm{vT}$ coming
from the r.h.s. of the triples equations with the corresponding
contributions in the SD(pvT) approximation. We found that the
individual terms agree at a reasonable 10\% level. The discrepancy
in the total value arises because there are certain very large
individual terms canceling each other. These terms are several
hundred times larger then the final combined result. In other words
there is a subtle cancelation taking place and our more
sophisticated all-order treatment profoundly affects this delicate
cancelation.

In addition, in Ref.~\cite{SafDerJoh98}, the  explicit contributions
of triples to the energies, $\delta E_\mathrm{vT}$, were computed
using direct third-order MBPT approach. Such terms are denoted in
Ref.~\cite{SafDerJoh98} as $E^{(3)}_{v,{\rm extra}}$, to emphasize
that these are diagrams missed in the SD approximation in the third
order. A comparison of our computed $\delta E_\mathrm{vT}$ with
$E^{(3)}_{v,{\rm extra}}$ is presented in Table~\ref{Tab:E3xtra}. We
again observe a large discrepancy, due to substantial cancelations
among contributions to $E^{(3)}_{v,{\rm extra}}$ and resulting
enhanced sensitivity to a correct all-order treatment.

The CCSD results obtained by \citet{EliKalIsh94} agrees with our CCSD energies
for the $3p_{1/2}$ state. However, for the $3s_{1/2}$ the two calculations disagree
by 14 cm$^{-1}$. This discrepancy is likely due to our omission
of all non-linear terms in the core CCSD equations.

\begin{table}[h]
\caption{Comparison of complementary third-order MBPT contributions
$E^{(3)}_{v,\,{\rm extra}}$ to removal energies with the corresponding
all-order correction $\delta E_\mathrm{vT}$.
The corrections are given in cm$^{-1}$. }
\label{Tab:E3xtra}
\begin{ruledtabular}
\begin{tabular}{lddd}
 &
 \multicolumn{1}{c}{$3s$} &
 \multicolumn{1}{c}{$3p_{1/2}$} & \multicolumn{1}{c}{$3p_{3/2}$}\\ \hline
$\delta E_\mathrm{vT}$                                 &  -25.4 &  -4.8 &  -4.7 \\
 $E^{(3)}_{v\,,{\rm extra}}$, Ref.~\cite{SafDerJoh98}  &   -9.2 &  -1.5 &  -1.6
\end{tabular}
\end{ruledtabular}
\end{table}

Comparing the final CCSDvT results for the removal energies with the
experimental values (last row of Table~\ref{Tab:E3xtra}) we find an
agreement at the level of 0.01-0.03\%. We do not include Breit-,
reduced-mass and mass-polarization corrections to the energies, as
they contribute at a much smaller level~\cite{SafDerJoh98}. A
perfect theory-experiment agreement for the previous SD(pvT)
calculations of energies~\cite{SafJohDer99} is fortuitous because
contributions of the disconnected excitations omitted in
Ref.~\cite{SafDerSaf99} would move the theoretical energies by about
70 cm$^{-1}$ for  the $3s_{1/2}$ state (see Table~\ref{Tab:Na_E}).

\subsection{Hyperfine constants and electric-dipole amplitudes}
\label{Sec:Mels}
With the computed wave functions of the $3s$,
$3p_{1/2}$ and $3p_{3/2}$ states we proceed to determining
magnetic-dipole hyperfine-structure constants $A$ and
electric-dipole transition amplitudes. The formalism was outlined in
Sec.~\ref{Sec:MelFormalism} and here we discuss our {\em ab initio}
results and compare them with the experimental values.

Numerical results are presented in Table~\ref{Tab:Na_MEs}. This
Table is organized as follows. First we list the DHF and SD values.
The results for the HFS constants include finite-nuclear size
effects (see Appendix). In the part denoted ``All-order corrections
beyond SD'' we tabulate differences between the values obtained at a
certain approximation (CCSD, SDvT, CCSDvT) and the corresponding SD
value (symbolically, e.g., $\Delta(\mathrm{CCSD}) = \mathrm{CCSD} -
\mathrm{SD}$). The most sophisticated approximation is CCSDvT (it
includes both implicit and explicit, Eq.~(\ref{Eq:ZT}),
contributions of valence triples and implicit contribution of
disconnected excitations); we will base our final {\em ab initio}
result on the CCSDvT values. A cursory look at this part of the
Table reveals that the contributions of disconnected excitations
tend to compensate contributions of valence triples for all the
computed properties. This situation is similar to the one observed
by us while presenting results for removal energies in
Section~\ref{Sec:Energies}.

While discussing the CCSDvT results, it is instructive to
compare the explicit valence
triple corrections to matrix elements, Eq.(\ref{Eq:ZT}), with
a corresponding contribution from the direct fourth-order
calculations~\cite{CanDer04}. In particular, for the
$\langle 3s ||D|| 3p_{1/2} \rangle$ amplitude,
the $Z_{wv}^{\left(T_v \right) }$  CCSDvT
contribution of -0.00075 is in a close
agreement with the fourth-order $Z_{1\times 2} (T_v)$ contribution
of -0.00073.
The close agrement is due to the fact that there are no strongly
canceling terms in the $Z_{1\times 2} (T_v)$ class of the fourth-order
diagrams. This should be contrasted with
our similar comparison of energy corrections (see Table~\ref{Tab:E3xtra}),
where large, a factor of 100, cancelations lead to a poor accuracy of
the direct third-order computation.

%

Corrections beyond  the CCSDvT approximation are listed in
the Table~\ref{Tab:Na_MEs} under the heading ``Complementary corrections''.
The dressing corrections arise due to a direct contribution of disconnected
excitations to the matrix elements. The details of our
all-order dressing scheme can be found in Ref.\cite{DerPor05}.
Following that work we distinguish  between vertex- and line-
dressing corrections. Futher, the ``MBPT-IV'' entries in the Table
include all  IVth diagrams missed by the CCSDvT method and dressing.
For example, our CCSDvT approximation discards core triples and disconnected
core excitations and these contributions arise starting from the  fourth order
of MBPT for matrix elements. In notation of Ref.~\cite{DerEmm02} the complementary
fourth-order terms are
$Z_{0\times 3}(D_v[T_c])$,
$Z_{0\times 3}(S_c[T_c])$, and
$Z_{1\times 2}(T_c)$. In addition,
the dressing method of Ref.\cite{DerPor05} misses so-called
stretched and ladder $Z_{1\times 2}(D_{nl})$ diagrams. These diagrams are
also incorporated into the value of ``MBPT-IV'' contribution in
the Table~\ref{Tab:Na_MEs}.
We used the fourth-order code of Ref.~\cite{CanDer04} to
evaluate the complementary MBPT-IV contributions.

Finally, we tabulate Breit and QED corrections available from the
literature (see Appendix for discussion). By combining them with the
CCSDvT values and the complementary corrections we arrive at the
final {\em ab initio} values in the bottom part of
Table~\ref{Tab:Na_MEs}. Here we also present a comparison with the
experimental data. In particular, the last row tabulates percentage
deviations from the experimental values. If the {\em ab initio}
value lays inside the experimental error bar, we tabulate
experimental uncertainty instead. The theory-experiment agreement is
better than 0.1\% except for the HFS constant of the $3p_{3/2}$
state, where our value disagrees with most accurate experimental
results at 1\% level. For this constant our result is, however, in a
reasonable agreement with the less accurate (0.3\% uncertainty)
result of Ref.~\cite{KriKusGau77}.

\begin{table}
\caption{Hyperfine structure
constants $A$ (in MHz) and  matrix elements of
electric dipole moment (in a.u.) for $^{23}$Na. Results of calculations and
comparison with experimental values. See text for the explanation
of entries. \label{Tab:Na_MEs} }
\begin{ruledtabular}
\begin{tabular}{lddddd}
 & \multicolumn{1}{c}{$ A(3s) $}     & \multicolumn{1}{c}{$ A(3p_{1/2})$}
 & \multicolumn{1}{c}{$A(3p_{3/2}$)}
 & \multicolumn{1}{c}{$\langle 3p_{1/2} ||D|| 3s \rangle$}
 & \multicolumn{1}{c}{$\langle 3p_{3/2} ||D|| 3s \rangle$}\\
 \hline
DHF                     &  623.8  &  63.39  &  12.59  &  3.6906   &  5.2188\\
SD                      &  889.0  &  95.05  &  18.85  &  3.5308   &  4.9932\\
\hline
\multicolumn{6}{c}{All-order corrections beyond SD}\\
$\Delta$(CCSD)           & -7.7   & -1.76  & -0.34  & 0.0072   & 0.0098 \\
$\Delta$(SDvT)          &     8.6 & 2.06   & 0.36      & -0.0115  & -0.0166 \\
$\Delta$(CCSDvT)        &     0.4 & 0.07  & -0.02   & -0.0035 &-0.0053 \\
\hline
\multicolumn{6}{c}{Complementary corrections }\\
Line dressing         & -2.4   & -0.43   & -0.09   &  0.0004    & 0.0005 \\
Vertex dressing       &  1.5   &  0.17   &  0.04   & -0.0001    &-0.0002 \\
MBPT-IV (core triples,...) & -2.8   & -0.41   & -0.06  & 0.0001   & 0.0001
\\
\hline
Breit + QED~\cite{SapChe03,SapChe05}   & 0.2     &       &  & 0.0001    & 0.0002    \\
\hline
Final CCSDvT + corrections\footnotemark[0]
                             &  885.9  &  94.45  &  18.72  & 3.5278  & 4.9885
\\
Experiment                  &  885.81 \footnotemark[1]
                                    &94.44(13) \footnotemark[2]
                                             &18.534(15)\footnotemark[3]
                                                        &3.5267 (17)\footnotemark[4]
                                                             &4.9875(24)\footnotemark[4] \\
                           &         &94.42(19) \footnotemark[5]
                                              &18.572(24)\footnotemark[6]
                                                        & 3.5246(23)\footnotemark[7]
                                                              &4.9839(34)\footnotemark[7] \\
                           &        &         &18.64(6)\footnotemark[8]& & \\
                           &        &         &18.69(9)\footnotemark[9]& & \\
Agreement with experiment & 0.01\% & <0.1\% & 1\% & <0.05\%  & <0.05\%
\end{tabular}
\end{ruledtabular}
\footnotemark[1]{Ref.~\cite{BecBokElk74}};
\footnotemark[2]{Ref.~\cite{WijLi94}};
\footnotemark[3]{Ref.~\cite{YeiSieHav93}};
\footnotemark[4]{Ref.~\cite{JonJulLet96}};
\footnotemark[5]{Ref.~\cite{CarJonStu92}};
\footnotemark[6]{Ref.~\cite{GanKarMar98}};
\footnotemark[7]{Ref.~\cite{VolSch96}};
\footnotemark[8]{Ref.~\cite{KriKusGau77}};
\footnotemark[9]{Ref.~\cite{AriIngVio77}}.
\end{table}

\section{Conclusion}
\label{Sec:concl}

To reiterate here we presented a practical high-accuracy {\em ab initio}
relativistic technique for calculating properties of  univalent atomic systems.
The distinct formal  improvements
over the previous singles-doubles
approach~\cite{BluJohLiu89,BluJohSap91,SafDerJoh98,SafJohDer99} are:
\begin{enumerate}
\item{non-perturbative treatment of valence triple excitations;}
\item{incorporation of disconnected excitations (non-linear terms) in
the coupled-cluster approach;}
\item{inclusion of complementary MBPT diagrams  so that the calculations
of matrix elements are complete through the fourth-order of MBPT; these diagrams
include contributions of core triples.}
\item{all-order ``dressing'' of lines and vertices in calculations
of matrix elements.}
\end{enumerate}
Including all the enumerated effects is important in reaching the
present uniform ``better than 0.1\%" theoretical accuracy for Na atom.
In particular,  a simultaneous treatment of triple and
disconnected quadruple excitations is required,
as these two relatively large effects tend to partially cancel
each other.

In the framework of the developed formalism, we computed removal
energies, magnetic-dipole HFS  constants $A$ and electric-dipole
amplitudes for the principal $3s_{1/2}-3p_{j}$ transitions. The
presented approach demonstrates a uniform sub-0.1\%-accurate
agreement with experimental data. In particular, we find that the
removal energies are reproduced within 0.01-0.03\%  and the HFS
constants of the $3s$ and $3p_{1/2}$ states with a better than 0.1\%
accuracy. The calculated dipole amplitudes are in a perfect
agreement with the 0.05\%-accurate experimental data. In the case of
the $3p_{3/2}$ state HFS constant our {\em ab initio} result
deviates from $\sim$ 0.1\%-accurate experimental
values~\cite{YeiSieHav93,GanKarMar98} by 1\%, while agreeing with
the less accurate measurements~\cite{KriKusGau77,AriIngVio77}. We
anticipate that the relativistic many-body technique presented here
can serve as a basis of highly-accurate evaluation of
parity-violating effects in Cs atom and Ba$^{+}$
ion~\cite{KoeSchNag03}.


\acknowledgements

We would like to thank Mark Havey for discussing results of
Ref.~\cite{YeiSieHav93}. This work was supported in part by the
National Science Foundation, by the NIST precision measurement
grant, and by the Russian Foundation for Basic Research under Grants
Nos.\ 04-02-16345-a and 05-02-16914-a.

\appendix
\section{ Smaller (non-correlation) corrections to the hyperfine structure constants}
Calculations of magnetic hyperfine constants $A$ presented in
Table~\ref{Tab:Na_MEs} were carried out with the nuclear
gyromagnetic ratio $g_I = 1.4784$. In calculations we model the
nucleus as a uniformly magnetized sphere of radius 3.83 fm. For the
$3s_{1/2}$ state, the corresponding nuclear size (Breit-Weisskopf)
effect reduces point-nucleus results by 0.5 MHz. In an extreme case,
when magnetization is assumed to be completely localized on the
nuclear surface, the $A_\mathrm{hfs}(3s_{1/2})$ is further reduced
by  0.15 MHz; this difference between the uniform and surface
magnetization is below our theoretical accuracy.

Breit and QED contributions to the HFS constant of the $3s_{1/2}$
state were calculated recently by \citet{SapChe03}. In their
notation, the value marked ``Breit/QED'' includes effects of the
Breit interaction, retardation in the transverse photon exchange and
negative-energy states, while ``QED'' correction encapsulates vacuum
polarization and self-energy corrections. (The Breit correction of
0.35 MHz,  evaluated using analytical expression~\cite{Sus01} is in
a reasonable agreement with the value of 0.2 MHz
from~\cite{SapChe03}). As to the QED corrections, the leading
Schwinger term (anomalous magnetic moment) $\delta A/A = \alpha/\pi$
sets a scale for radiative corrections at 0.1\% and this is
comparable with the accuracy of our calculations. Nevertheless,
explicit model-potential calculation~\cite{SapChe03} of vacuum
polarization and self-energy corrections displays a large degree of
cancelation between different contributions, leading to the total
QED correction 70 times smaller than the Schwinger term.

Following discussion of Ref.~\cite{YanMcKDra96} for Li, we also
analyzed the following smaller corrections to the HFS constant: (i)
Mass scaling. This effect contributes at the relative level of $
1/(1 +m_e/M_{\rm nuc})^3 \approx 7 \times 10^{-3}$ \%; here $M_{\rm
nuc}$ is the nuclear mass. (ii) Mass polarization. It occurs due to
an additional introduction of the term $- \mu/M_\mathrm{nuc}
\sum_{i>j} \mathbf{\nabla}_i \cdot \mathbf{\nabla}_j$ into the
atomic Hamiltonian, $\mu$ being the reduced mass of the electron. We
expect that this term would contribute at the relative level of
$1/M_\mathrm{nuc} (\alpha Z)^2 \approx  10^{-5}$ \%. (iii) Second
order in magnetic-dipole HFS interaction. It contributes at the
$10^{-5}$ \% level. All the enumerated corrections are below the
level of theoretical accuracy of the calculation presented here.



\end{document}